\documentstyle[12pt]{article}

\begin{document}

\begin{titlepage}
August, 1992 \hfill                 UTHEP-242
\begin{center}

{\LARGE\bf A Study  of the $N=2$ Kazakov-Migdal Model}

\vspace{1.5cm}
\large{
Sinya Aoki\\
Institute of Physics, University of Tsukuba\\
Tsukuba, Ibaraki 305, Japan\footnote{Permanent address} \\
and \\
Physics Department, Brookhaven National Laboratory\\
Upton, NY 11973, USA \\}
\vspace{0.5cm}
\large{
and\\
}
\vspace{0.5cm}
\large{
Andreas Gocksch and Yue Shen\\
Physics Department, Brookhaven National Laboratory\\
Upton, NY 11973, USA \\
}
\vspace{1.9cm}
\end{center}

\abstract{
We study numerically the SU(2) Kazakov-Migdal model of `induced QCD'.
In contrast to our earlier work on the subject we have chosen here {\it not}
to integrate out the gauge fields but to keep them in the Monte Carlo
simulation.
This allows us to measure  observables associated with the
gauge fields and thereby address the problem of the local $Z_2$ symmetry
present
in the model. We confirm our previous result that the model has a line of
first order phase transitions terminating in a critical point. The adjoint
plaquette has a clear discontinuity across the phase transition, whereas
the plaquette in the fundamental representation is always zero in accordance
with Elitzur's theorem.
The density of small $Z_2$ monopoles shows very little variation and
is always large.
We also find that the model has extra local U(1) symmetries
which do not exist in the case of the standard adjoint theory.
As a result, we are able to show that two of the angles
parameterizing the gauge field completely decouple from the
theory and the continuum limit defined around the critical point can
therefore not be `QCD'.
}
\vfill
\end{titlepage}

\section*{Introduction}
Recently Kazakov and Migdal (KM) \cite{KM} presented an entirely new approach
to the problem of finding a solution to QCD in the large $N$ limit. In a
nutshell,
their philosophy is as follows: Suppose there is substructure to the quarks and
gluons which becomes visible at energies large compared to some scale
$\Lambda$.
Then, regardless of the detailed nature of the constituents one should,
upon integrating them out
end up with QCD at the scale $\Lambda$ as the only renormalizable,
asymptotically free theory
in four dimensions.  KM propose to use a scalar field in the
adjoint representation as a gluon constituent.
The scalar mass acts as an effective ultraviolet cutoff for
the theory. KM
consider the following model, defined on a d-dimensional hypercubic
lattice with action
\begin{equation}
S = N \sum_x \left[ V(\Phi(x))-tr \sum_{\mu = 1}^d \Phi(x)U_\mu(x)
\Phi(x+\mu)U^\dagger_\mu(x)\right]~,
\end{equation}
where $\Phi(x)$ is a traceless scalar field in the adjoint
representation of $SU(N)$, covariantly coupled to the gauge fields
$U_\mu(x)$ defined on the links of the lattice.
At distances on the order of the lattice spacing this theory clearly
looks very different from ordinary QCD defined on the lattice. But perhaps,
KM argue, if we can find a critical point of the lattice theory at which
the continuum limit can be taken and if furthermore we can manage to keep
the scalar field heavy in that limit then QCD might be `induced'.

Let us see how the above scenario could possibly work in the limit of
smooth weak gauge fields.
Integrating out the scalar fields in the partition function will give
rise to an induced action for the gauge fields which because of the
gauge invariance of Eq. (1), will be given as an (infinite) sum over Wilson
loops on the lattice. Moreover, because the scalar field is in the
adjoint representation, the Wilson loops will also be. In the case of
$V(\phi) = \frac {m_0^2}{2}tr{\phi}^2$, the $\phi$ integral is Gaussian
and one obtains the induced action \cite{KM}:
\begin{equation}
S_{ind} = - \sum_{\Gamma} \frac {2^{l(\Gamma)-1}|trU(\Gamma)|^2}
{l[\Gamma]m_0^{2l(\Gamma)}}~,
\end{equation}
where $l(\Gamma)$ is the length of the path $\Gamma$.
To leading order this model has a critical point at
$m_0^2 = 2d$ as can be seen easily from Eq. (1) where the lowest order
action becomes an action for a free scalar field. Expanding in fluctuations
a $F_{\mu\nu}^2$ term is generated and
using the definition $m^2 = m_0^2 - m_c^2$ KM \cite{KM}
obtained the correspondence
\begin{equation}
\frac{1}{g_0^2} \rightarrow -{N\over 96\pi^2} ln(m^2a^2) ~.
\label{eq:bareg}
\end{equation}
This result arises simply from the scalar loop.
In the scaling region this model is supposed to be QCD with
$m$ acting as an {\it ultraviolet} cutoff.
Denoting by $M_g$ a physical
glueball mass this immediately gives the usual relation between a physical
quantity
, the cutoff and the bare coupling
\begin{equation}
ln \frac{m^2 a^2}{M_g^2 a^2} = {48\pi^2 \over 11N g_0^2}~.
\end{equation}
Combining this expression with that of Eq. (\ref{eq:bareg}) one gets
a power law scaling
\begin{equation}
M_g^2 =  (m^2)^{\gamma}~,
\end{equation}
with $\gamma = 23/22$!

If we assume that the above arguments are in fact correct then the possible
payoff of the particular choice of inducing field is indeed enormous:
Based on the
work of KM and of Migdal \cite{migdal} it looks as though in the limit
of large $N$ the model is  analytically tractable and that the degrees
of freedom associated with the eigenvalues of
$\Phi$ could actually be the long sought ``master field" of large $N$
QCD.
The model is easier dealt with analytically and numerically if one
{\it first} integrates out the gauge fields.
Migdal \cite{migdal} has found scaling solutions to the equation
for the eigenvalue distribution in the
model and obtained
$\gamma = 1.365..$ in four dimensions. More recently he even succeeded
in obtaining the spectrum of small fluctuations around these
solutions\cite{migdal2}.

To summarize this section: If the model defined by Eq. (1) is to induce
QCD correctly we expect the system can be tuned to a critical point and
power-law scaling of physical quantities
with non-trivial (non mean field) critical exponents.

\section*{The $SU(2)$ Phase Diagram }
The road to proving that the model in Eq. (1) induces QCD is a long one.
First one must demonstrate that the model for arbitrary $N$ has indeed
a critical point at which a continuum limit can be constructed.
This is not a trivial point: Lattice Higgs models typically have first order
phase transitions and critical behavior is more an exception than the rule.
The KM model is actually not quite
unknown to lattice theorists - in the case of SU(2) it is the lattice version
of the
Georgi-Glashow \cite{georgi} model in the limit of infinite
gauge coupling.
This model has been studied before \cite{latgg} but
the phase structure in the case of infinite gauge coupling
was until recently unknown. Once the existence of a critical
point has been established one
should (ideally) calculate the spectrum and check whether (5)
is satisfied. Finally, in order for an
exact solution of the theory at large $N$ to be relevant for
physics at $N=3$ one must show that there are no unexpected phase
transitions in $N$. As to the last point, the K-M model has recently been
solved
in one dimension at arbitrary $N$ and it was found that the $N$ dependence
is smooth \cite{italians}. However in 1-D the model is trivial unless one
compactifies
the line to a circle and even then it is effectively a one point model
since one can always go to a gauge in which all the links but one
are equal to unity. This seems like a rather severe over-simplification.

In the following we will study $N=2$ case which is
the simplest non-trivial version of
the model. Our objective is to determine whether it is possible to
construct a continuum limit for the theory and in what sense
it could be equivalent to the $SU(2)$ gauge theory (`QCD') in that limit.
First we must search for a critical point. In our previous work \cite{shen}
we did just that by integrating out the gauge fields following
KM's suggestion. One obtains
\begin{equation}
Z = \int_{\phi >0}[d\phi]
\exp \left\{\sum_x \left[ ln{\phi}^2-2V(\phi)\right]+\sum_{<xy>}
ln\left[{sinh(4\phi(x)\phi(y))\over \phi(x)\phi(y)}\right]\right \}~,
\label{eq:scalar}
\end{equation}
where $<xy>$ denotes the sum over nearest neighbors of x.
The field $\phi(x)$ in Eq. (\ref{eq:scalar}) is the
(positive) eigenvalue of the matrix $\Phi$.
In this form the problem
is to determine whether the above scalar theory has a nontrivial continuum
limit.
It is well known that the continuum limit of
a scalar model with polynomial interaction terms is trivial with vanishing
renormalized coupling \cite{wilson}.
In principle the continuum limit of
Eq. (\ref{eq:scalar}) could avoid the triviality problem
due to the presence of the nonpolynomial interaction terms.
But note that these terms appeared as a consequence of the fact that
we could integrate out the phase of the scalar by doing the
$U-$ integral exactly.
If we could integrate out the angular variables in an $O(N)$ model exactly
the resulting effective action for the radial variables would also be
nonpolynomial. The phase diagram that was found in Ref. \cite{shen},
corresponding to
a scalar potential of the form $V(\phi) = \frac
{m_0^2}{2}tr\phi^2 + \frac{\lambda}{4}tr\phi^4$, is shown in Fig. 1.
There is a first order phase transition at small $\lambda$ which ends
in a critical point. In a simple mean field theory the critical value
 is $\lambda =2.57, m_0^2 = 4.52$. From the Monte Carlo the precise location
of the endpoint is very hard to determine. But it's existence is
fortunate: It allows us to construct a continuum limit.
Interestingly though,
in the $\lambda \to 0$ limit
{\it no} continuum limit exists. This is consistent with the observations
by Gross \cite{gross} who constructed an exact large $N$ solution in
the case of a quadratic potential.

\section*{The Nature of the Phase Transition}
In this section we will present the results of a Monte Carlo study
of the theory as described by the action in Eq. (1). The gauge
invariant content of the $\Phi$- field lies in it's eigenvalues,
so we again diagonalize $\Phi$ and write
\begin{equation}
Z =\int [dU] \int_{\phi >0}[d\phi]
\exp \left\{\sum_x \left[ ln{\phi}^2-2V(\phi)\right]
+4 \sum_{\mu = 1}^d \phi(x)
\phi(x+\mu)(2 |{z_\mu^0(x)}|^2-1)\right]~.
\label{eq:scalar1}
\end{equation}
Here we have made use of the parameterization
$U=\left(\matrix{z^0&z^1\cr-\bar{z}^1&{\bar z}^0\cr}\right)$.
Keeping the gauge fields in place will allow us to study the nature of the
phase transition
that was discussed in the previous section.

Note that the theory has in addition to the local $SU(2)$ gauge invariance
additional {\it local} $U(1)$ symmetries.
This very important fact was pointed out by Gross \cite{gross}.
The center symmetry $U_\mu(x) \rightarrow
Z_\mu(x)U_\mu(x)$
discussed by Kogan, Semenoff and
Weiss \cite{kogan}, which is responsible for the fact that only adjoint loops
appear in Eq. (2),
is a subset of this larger symmetry.
We can make the larger symmetry very explicit by using ordinary gauge
invariance
to diagonalize the $\Phi$ field as in (7) to obtain the gauge field action
(in unitary gauge')
\begin{equation}
S_G=
2 \sum_{x,\mu = 1}^d \phi(x)
\phi(x+\mu)tr[\tau_3 U_{\mu}(x) \tau_3 {U_{\mu}}^\dagger(x)]
\end{equation}
The remaining symmetry now after the diagonalization
is ($V,W$ are diagonal matrices $\in SU(2)$)
$U_\mu(x) \rightarrow V_\mu(x)U_\mu(x){W_{\mu}}^{\dagger}(x)$.
Since one of them is the diagonal part of the usual $SU(2)$ gauge
transformation, we have $d\times 2 - 1$ extra $U(1)$ symmetries
\footnote{It is hard to see this symmetry in the induced action, eq.(2).}.
Now use the following (Euler) parameterization of the links
\begin{equation}
U=\left(\matrix{e^{i \theta_1}&0\cr0&e^{-i
\theta_1}\cr}\right)\left(\matrix{\sqrt{1-b^2}&b\cr-b&\sqrt{1-b^2}\cr}\right)
\left(\matrix{e^{i \theta_2}&0\cr0&e^{-i \theta_2}\cr}\right).
\end{equation}
In terms of these parameters we obtain
\begin{equation}
S_G=4 \sum_{x,\mu = 1}^d \phi(x)
\phi(x+\mu)(2 {b_{\mu}}^2(x)-1).
\label{eq:eq10}
\end{equation}
Hence the reason for why (7) depends on only one real parameter associated with
the links is apparent: When the $\Phi$ field is diagonal the additional
$U(1)$ symmetries can be used to eliminate two (diagonal) degrees
of freedom from each link. Eq. (10) does not depend on the $\theta$ angles in
Eq. (9)\footnote{In $SU(3)$ we also found a parameterization in which
$U=D_1 V D_2$ ($D_1$, $D_2$ are diagonal matrices $\in SU(3)$) so that
the action in this case only depends on 4 parameters. We suspect that this
generalizes
to $SU(N)$: the action only depends on $(N-1)^2$ parameters.}.
Note that our counting differs by a factor of 2 from that of Gross,
which apparently was based on counting degrees of freedom in the naive
continuum limit.
It is indeed true that in the naive continuum limit where $U_{\mu}=1+iA_{\mu}$
only the diagonal component $A_3$ drops out of the action. However as we can
see
from Eq. (10) the action depends only on {\it one} component of the gauge
field.
This is important: The naive continuum limit is in fact too naive in the
KM model. The correct limit is obtained by writing the $b$ dependent matrix
in Eq. (8) as $e^{iA \tau_2}$ and expanding in $A$. The resulting action
is (10) with $b$ replaced by $A$. The significance of this observation
lies in the fact that in the weak coupling (i.e. large $\phi$) limit
only the $b_{\mu}$ degree of freedom becomes smooth, the two angles
$\theta_1, \theta_2$ in Eq. (10) are always randomly distributed.
The importance of the correct continuum limit was recently pointed out
in a new publication of Kogan et al. \cite{kogan2}.

In this paper we have monitored in addition to the $\phi$ field also
the expectation value of the plaquette in the adjoint representation
\begin{equation}
P_{adj}={1 \over 3}(|trU(P)|^2-1).
\end{equation}
Due to the local $Z_2$ symmetry of the model Wilson loops in the
fundamental representation vanish according to Elitzur's theorem
\cite{elitzur}.
Based on the behavior of this quantity in addition to that of the
`density of $Z_2$ monopoles'
\begin{equation}
\bar \rho=<1- \prod_{P \in C_3}sign(trU(P))>
\end{equation}
we will later on speculate on the significance of the $Z_2$ degrees
of freedom in the KM model. Eq. (9) and (10) imply that we can derive a simple
expression for $P_{adj}$. The Haar measure for the parameterization (9)
is
\begin{equation}
dU=2b db d\theta_1 d\theta_2.
\end{equation}
Using the fact that $b$ enters the action
only quadratically we can completely integrate out the gauge field to
obtain a formula for the adjoint plaquette in terms of $\phi$. Denoting by
\begin{equation}
C=\left( \matrix{1-a&a\cr
a&1-a\cr} \right),
\end{equation}
where
\begin{equation}
a={1 \over 8 \omega}-{1 \over e^{8 \omega}-1},
\end{equation}
a matrix on the links\footnote{ In general $(U^\dagger)_{ij} U_{kl}
\rightarrow \delta_{il}\delta_{jk} (C)_{ij}$ .}
 which depends
on $\phi$ at the beginning and the end of the link through $\omega
=\phi(x)\phi(x+\mu)$
we obtain
\begin{equation}
<P_{adj}>={1 \over 3}<(tr(C_1C_2C_3C_4)-1)>.
\end{equation}
The average is over the $\phi$ field.
In mean field theory, where we set $\phi =const$ and determine
the mean field from the minimum of the resulting potential one obtains the
simple result
\begin{equation}
<P_{adj}>={1 \over 3}(1-2a_{MF})^4.
\end{equation}
Later this will be compared to the result of our simulation.
Note one important consequence of this result: The normalization of $P_{adj}$
is such that in the naive continuum limit $P_{adj}=1$. This is obviously
{\it wrong} though. From Eq.(14) we see that in the weak coupling limit
\footnote{
In the strong coupling limit($\omega \rightarrow 0$) $P_{adj} \rightarrow 0$.}
($\omega \rightarrow \infty$) $P_{adj} \rightarrow {1 \over 3}$.
Again this agrees with the observations of Ref. \cite{kogan2}.
It is also easy to see that the angular integrations imply that $\bar \rho =1$.

We have simulated the theory described by
the action in Eq. (\ref{eq:scalar1}) on both $4^4$ and $8^4$ lattices using
a simple Metropolis algorithm on both the links and $u$ where $\phi=e^u$.
It turns out that for the local quantities we measure there is hardly any
difference
in the results on the two lattices. Of course this is consistent with the fact
that the transition is first order. As far as the the $\phi$ expectation value
is
concerned we reproduce the results of Ref. \cite{shen} where the gauge fields
were integrated out exactly. This is a good check on the program. In Fig. 2 and
3 we show
a thermal cycle for $\phi$ and
the adjoint plaquette, $<{1 \over 3}(|trU(P)|^2-1)$.
There is a clear discontinuity in this
quantity just as one would expect based on simple mean field theory. Clearly
the agreement between Monte Carlo and mean field theory is excellent.
The adjoint plaquette close to the
critical point is shown in Fig. 4. The magnitude of the gap (if there is any)
has decreased considerably. Also note that the adjoint plaquette rises only
very
slowly. At small $m_0^2$ it is only $O(0.14)$ which is far away from the weak
coupling
limit $P_{adj} = {1 \over 3}$. Hence relatively close to the end point of the
first order
phase transition the gauge field is far from being perturbative!

We have also measured the fundamental plaquette
in various places in the phase diagram both above and below the phase
transition line
in Fig. 1. It is always zero. An interesting quantity is the {\it local}
plaquette which is histogramed in Figures 5 and 6. Note how in the weak
coupling
(small $m_0^2$) phase peaks develop at ${1\over2}trU =1,-1$. One might think at
first that this
has something to do with some kind of `$Z_2$ Higgs' transition advocated in
Ref. \cite{kogan}. However the explanation is much simpler than that:
At weak coupling, when $b \rightarrow 0$, the local plaquette is roughly
given by
\begin{equation}
{1 \over 2}trU(P) \approx cos(\theta_P)
\end{equation}
where $\theta_P=(\theta_1+\theta_2)_P$.
Averaging over $\theta_P$ with uniform distribution gives a vanishing
plaquette but the (normalized) distribution will be approximately
\begin{equation}
W({1\over 2}trU(P)) \approx {1 \over 2 \pi}{1 \over \sqrt{1-({1 \over
2}trU)^2}}.
\end{equation}

We have also measured $\bar \rho$.
This quantity is always large, $\bar \rho \approx 0.96$ showing hardly any
variation
as one moves around in the phase diagram. As we mentioned before based on the
randomness of the two angles the expected value of this quantity is unity.
In the next section we will discuss the significance of this observation.

\section*{Discussion and Conclusion}
In light of Eq. (2) one is tempted to view
the KM model as a theory with a generalized action in the adjoint
representation.
The novel feature here is that there is actually
an infinite sum over arbitrary shaped loops. In the naive
continuum limit this action seems to be as good as any other in that it
produces
a term proportional to ${F_{\mu \nu}}^2$, with a non Abelian field strength $F$
(the Abelian component drops out). But as we saw in the last section
the naive continuum limit is incorrect in this model and we must be careful
here.

What is the mechanism that drives the phase transition?
At first look the KM model reminds us of
the studies of the $SO(3)$ invariant theory 10 years ago.
It was found that the mixed action theories have first order phase transitions
in the ($\beta_F$,$\beta_A$) plane, including the point $\beta_F =0$
\cite{mike}.
An elegant explanation of why one expects a first order phase transition
in the $SO(3)$ theory was given by Halliday and Schwimmer \cite{schwimmer}:
The theory contains zero energy monopoles (in contrast to the SU(2) case
where the presents of monopoles costs energy) which condense at some
critical value of the coupling. In the monopole free phase the adjoint
and fundamental theory approach a common continuum limit.
Halliday and Schwimmer base their considerations on the following
$SO(3)$ invariant theory:
\begin{equation}
S=\beta \sum_P \sigma(P) trU(P)
\label{eq:mono}
\end{equation}
where the sum is over all plaquettes on the lattice. The variable $\sigma$
takes
on values in $Z_2$ and lives on plaquettes. Under $Z_2$ it transforms as
$\sigma \rightarrow (\prod_{l \in P}z_l) \sigma$.
A string of plaquettes carrying a nontrivial $\sigma$ can be thought of as
the Dirac string of a monopole. The location of the monopole can be identified
with an end of the string defined as a point on the lattice where
\begin{equation}
\rho =\prod_{P \in C_3}\sigma(P)
\end{equation}
is negative. Here $C_3$ is a three dimensional cube on the lattice.
Halliday and Schwimmer argue
that in the phase in which the monopoles are absent we can write
$\sigma \approx \prod_{l \in P}z_l $ in which case the $z_l$ can be absorbed
into
the links. In particular for a `tiled' Wilson loop we have
\begin{equation}
<\prod_{l \in C} U(l) \prod_{P \in A(C)} \sigma(P)>_U \approx <\prod_{l \in C}
U^{\prime} (l)>_{U^\prime}
\label{eq:mono2}
\end{equation}
with $U^{\prime} (l) = z_l U(l)$. The above equation tells us that `tiled'
(i.e. $Z_2$ invariant)
Wilson loops in the adjoint theory described by Eq. (\ref{eq:mono2}) will
behave like
ordinary Wilson loops in the theory defined using the fundamental
representation. Note, that
this does {\it not} mean that the fundamental Wilson loop is ever non zero in
the
adjoint theory. This is not allowed as a consequence of Elitzur's theorem
\cite{elitzur}.
For this reason we do not quite understand the analysis in Ref. \cite{kogan}
where it is claimed that there can be a `Higgs' phase with a nonvanishing
fundamental plaquette.
Monopole condensation can be checked
by computing the monopole density $<\bar \rho>$
which is invariant and always non-zero. Monte Carlo simulations \cite{mc}
show a sharp rise of this quantity across the transition making the case for
the above scenario of the phase transition quite convincing.

Does this scenario apply to the $SU(2)$ KM-model? We do not think so.
We have seen that
the quantity $\bar \rho$ is always large and shows absolutely no variation
as we go through the phase transition. Hence $Z_2$ monopoles are always
condensed in our system! This is of course due to the fact that as far
as the $\theta$ degrees of freedom are concerned the theory is {\it always}
in the strong coupling phase. We therefore conclude that the first
order phase transition is {\it not} a $Z_2$ driven transition.
If one wanted to understand
the phase transition entirely based on the formulation in terms of
adjoint loops one must keep all the loops and probably introduce
`large $Z_2$ monopoles' for all of them. These `monopoles' will interact and
the picture will become a hopeless mess. Of course somehow this `covariant'
view (as opposed to `unitary' one based on diagonalizing $\Phi$)
of the phase structure must be equivalent to the much simpler
description of the phase transition in terms of Eq. (10).

In the description of Eq. (\ref{eq:eq10}) the $Z_2$ degrees of freedom have
completely decoupled and are just part of the randomly fluctuating $\theta$
variables. The transition is described very well by simple
mean field theory which `explains' the transition in the
same way we explain for example the liquid-gas transition.
We have seen that
the one relevant degree of freedom of the gauge field becomes perturbative
at small $\lambda$ (in the weak coupling phase).
Close to the end point of the transition line
the adjoint plaquette is still rather small. This is of course due to the
fact that $\phi$ is small there.
Assuming for the moment that by carefully tuning
the parameters of the scalar potential we can construct a continuum
theory, what will it be like? From what we have seen it certainly
does not look like the $SU(2)$ gauge theory. Since two internal degrees of
freedom have disappeared per component of the gauge field the most
likely scenario is that we will just get a (trivial?) theory
involving a massive photon (with two transverse and a longitudinal components)
and a scalar.

For large $N$
things could be different: There the theory retains it's non Abelian nature
even after all symmetries have been used to eliminate degrees of freedom.
Even in this case the question remains in what sense the theory becomes
equivalent to QCD at the critical point. One would for example like
to establish that there is a string tension in the continuum limit.
Fundamental Wilson loops always vanish in the KM-model whereas adjoint
loops have a perimeter law due to screening. Maybe the answer are the
`tiled loops' (Wilson loops whose minimal area
is tiled with plaquettes) introduced by Kogan et al. \cite{kogan2} although
they have the strange property of not vanishing at infinite coupling.
Migdal in the meantime \cite{migdal3} has proposed to discard the simplest
version of induced QCD and instead add in addition to the adjoint scalar
also $n_f$ ($1 << n_f << N$) fundamental fermions to the action. In this
approach
the fundamental Wilson loop need not vanish and may contain in addition to
the perimeter piece also an area law contribution. We have also just become
aware
of a new paper by Kogan et al. \cite{kogan3} where it is argued in mean field
theory
($SU(2)$) that critical behavior could be recovered in the case $\lambda=0$ by
changing the measure for the scalar. While this might be true, based on what
we have found the resulting theory will not be `QCD'.

\section*{Acknowledgements}

We thank M. Creutz for valuable discussions.
S. A. thanks members of the theory group at Brookhaven National Laboratory
for their kind hospitality during his stay.
This work was supported by a DOE grant at
Brookhaven National Laboratory (DE-AC02-76CH00016). The numerical
simulations are performed at SCRI at Tallahassee.

\pagebreak

\section*{Figure Caption}

\noindent {\bf Figure 1}: The phase diagram. The solid line is the mean
field theory prediction and the simulation results are indicated by diamonds.
The end point of the first order phase transition line is marked by *
for MFT and square for Monte Carlo estimate. At the end point the theory
becomes critical.

\noindent {\bf Figure 2}: Comparison of the mean field theory and
Monte Carlo simulation result for $\phi$-field vacuum expectation value.
The solid lines are the MFT predictions and
Monte Carlo data points are indicated by *. The errors are smaller than the
size of the symbols.

\noindent {\bf Figure 3}: Comparison of the mean field theory and
Monte Carlo simulation result for the adjoint plaquette.
The solid lines are the MFT predictions and
Monte Carlo data points are indicated by *.

\noindent {\bf Figure 4}: Monte Carlo simulation result for the adjoint
plaquette.
The value of $\lambda$ is close to the critical value.

\noindent {\bf Figure 5}: Histogram of the local plaquette in the
weak coupling phase.

\noindent {\bf Figure 6}: Histogram of the local plaquette in the
strong coupling phase.

\pagebreak

\end{document}